\documentclass[11pt,a4paper,onecolumn]{scrartcl}
\usepackage[utf8x]{inputenc}
\usepackage[english]{babel}
\usepackage{amsmath}
\usepackage{amsfonts}
\usepackage{amssymb}
\usepackage{mathtools}
\usepackage{graphicx}
\usepackage{threeparttable}
\usepackage[table.xcdraw]{xcolor}

\author{Philipp Koch\footnote{~Correspondence: \href{mailto:philipp.koch@ecoaustria.ac.at}{philipp.koch@ecoaustria.ac.at}}~$^,$\footnote{~EcoAustria - Institute for Economic Research, Am Heumarkt 10, 1030 Vienna, Austria.}~, Clemens Fessler\footnote{~Correspondence: \href{mailto:h11842538@wu.ac.at}{h11842538@wu.ac.at}}~$^,$\footnote{~Vienna University of Business and Economics, 1020 Vienna, Austria.}}
\title{A test for Heckscher-Ohlin using value-added exports\footnote{~We would like to thank Harald Badinger and Ingrid Kubin for their valuable comments.}}
\subtitle{}
\usepackage{fancyhdr}
\usepackage{setspace}
\usepackage[olditem,oldenum]{paralist} 
\usepackage{caption}
\usepackage{subcaption}
\usepackage{remreset}
\usepackage{pdfpages}
\usepackage[hidelinks]{hyperref}
\usepackage{longtable}
\usepackage{tabularx}
\usepackage{chngcntr}
\usepackage{amsthm}
\usepackage{adjustbox}

\makeatletter
\@removefromreset{equation}{chapter}
\makeatother

\usepackage{apacite}

\begin{document}

\maketitle

\begin{abstract}
\begin{center}
    \sffamily\bfseries{Abstract} 
\end{center}
Empirical evidence for the Heckscher-Ohlin model has been inconclusive. We test whether the predictions of the Heckscher-Ohlin Theorem with respect to labor and capital find support in value-added trade. Defining labor-capital intensities and endowments as the ratio of hours worked to the nominal capital stock, we find evidence against Heckscher-Ohlin. However, taking the ratio of total factor compensations, and thus accounting for differences in technologies, we find strong support for it. That is, labor-abundant countries tend to export value-added in goods of labor-intensive industries. Moreover, differentiating between broad industries, we find support for nine out of twelve industries. 
  \par
\end{abstract}

\section{Introduction}
Empirical evidence for the predictions of the Heckscher-Ohlin model is mixed. While some studies support the reasoning that countries export relatively more of the goods which use their abundant production factor intensively \cite{Romalis, Morrow}, others do not find results in favor of the Heckscher-Ohlin Theorem \cite{Baldwin, Trefler}. One potential reason for this shortcoming is that data on gross trade is used, while the predictions of trade models may be more accurate for trade in terms of value-added, as \citeA{Leamer} already pointed out. \par
With the recent availability of data on trade in value-added, the aim of this paper is to investigate whether the Heckscher-Ohlin Theorem accurately predicts value-added trade flows. Methodologically, we employ a similar approach to \citeA{Ito.2017}, who were among the first ones to empirically test the Heckscher-Ohlin framework with value-added trade data. Specifically, they use the 2013 release of the World Input-Output Database \cite{Timmer.2015} and show that skill-abundant countries tend to export value-added in products of skill-intensive industries. However, they do not find evidence that labor-(capital-)abundant countries tend to export value-added in products of labor-(capital-)intensive industries. \par
Since more detailed data on capital and labor is available and the modelling of the Socioeconomic Accounts has been improved with the 2016 release of the WIOD \cite<see>{Gouma.2018}, we aim to reinvestigate this issue, and test whether the Heckscher-Ohlin Theorem with respect to labor and capital holds for trade in value-added. \par
The paper is structured as follows: Section 2 provides a short description of the Heckscher-Ohlin framework. Section 3 reviews the empirical evidence. Section 4 describes the applied data sources in this paper, and the calculation of value-added exports. Then, we describe the methodological approach, before we report our results and conclude.

\section{The Heckscher-Ohlin Model and its empirical evidence}
Although foundations of the model were already laid out earlier in his dissertation, the main theories of the Heckscher-Ohlin model were presented in a book published by the Swedish economist Bertil \citeA{Ohlin}. However, Eli \citeA{Heckscher} had studied the topic as well and some of the elements in Ohlin’s work came from an article published by Heckscher in 1919, which is why he was mentioned as a co-developer. \par
The Heckscher-Ohlin Theorem, one of the central theorems in the model, links relative factor endowments to patterns of international trade. Assuming that both countries have the same technology but differ in their initial factor endowments, the Theorem states that countries export goods, which use their abundant factor more intensively. That is, we should be able to observe that countries with an abundance in labor export goods using labor more intensively. \par

\subsection{Empirical evidence}
Empirical evidence on the predictions of the Heckscher-Ohlin Theorem is mixed. Most early empirical studies on the Heckscher-Ohlin Theorem were carried out on US data. \citeA{Leontief} used US trade data to test the model’s predictions. He found that, even though the US have an abundance of capital, they export relatively more labor-intensive goods, while also importing more goods that were relatively more capital-intensive. These results go strictly against the predictions of the Theorem and became known as the Leontief Paradox. \par
Multiple studies have since been carried out with the goal of testing the Heckscher-Ohlin Theorem in different settings to find an explanation for the Leontief Paradox. \citeA{Bowen.1987} employed an extensive cross-sectional data set including 27 countries and 12 factors. Carrying out sign tests, they were able to support the model’s prediction in only 61\% of all cases. Other early empirical tests, such as the papers written by \citeA{Baldwin} and \citeA{Leamer} were also unable to confirm the Theorem. \par
One possible reason for this mixed evidence is that some of the strong assumptions that are necessary for the model  may not hold in reality. In particular, it is doubtful that technologies are in fact equal across countries. One indication that this is indeed not the case was presented by \citeA{Trefler}. He proposes that differences in a country’s factor abundances should allow conclusions about its trade volumes. However, when testing the hypothesis, he finds that trade volumes are substantially lower than what the model would predict. \citeA{Trefler} then shows that the predictions become significantly better once differences in technologies, and therefore in productivities, across countries are being accounted for. Following this approach, \citeA{Weinstein} test the Heckscher-Ohlin Theorem while relaxing the assumption of common technologies on a data set including 10 countries. They find that in this framework the predictive power of the model increases both when performing a sign test on the direction of trade as well as when looking at missing trade. However, differences in predictive power vary strongly and depend on the exact specifications of the model. \par
\citeA{Romalis} chooses a different approach and introduces a model with transport costs and monopolistic competition. When he tests the model’s predictions using bilateral trade data for the United States, he finds that countries trade those commodities more strongly that use their abundant factors more intensively. Extensions of the model by \citeA{Morrow} and \citeA{Regolo} are also able to support the HO-Theorem in most specifications with their respective samples. \par
The papers mentioned so far show that verifying the predictions of the Heckscher-Ohlin Theorem is difficult and depends on both the framework used and the data available. One further explanation is that data on gross exports may not adequately capture what countries are in fact producing competitively.\footnote{~For example, \citeA{Timmer.2019} show that export specializations differ depending on whether one investigates gross or value-added trade.} Specifically, gross exports do not only incorporate domestic value-added but also foreign value-added and double-counted exports \cite{Koopman.2014}.  \par
The difference between gross and value-added trade data is becoming increasingly more relevant, as global value chains have been deepening over the last decades \cite{Johnson.2017}. It was already pointed out by \citeA{Leamer} that using value-added trade data may be necessary to provide evidence for Heckscher-Ohlin. However, there has been no data to adequately test whether trade in value-added follows the predictions more closely. The release of the World Input-Output Database (WIOD) in 2010 and other inter-country Input-Output Tables have enabled such testing and sparked an increase in publications working with the data set. One recent and relevant study by \citeA{Ito.2017} uses the WIOD in order to investigate whether the explanatory power of Heckscher-Ohlin differs depending on whether gross export or value-added data is employed. \par 
The data used by \citeA{Ito.2017} spans from 1995 to 2009 and includes 40 countries with 34 sectors. They follow the approach by \citeA{Chor}, thereby generalizing the approaches developed by \citeA{Weinstein} and \citeA{Trefler.2010}. When looking at manufacturing, \citeA{Ito.2017} find that countries from their sample, which are relatively skill-abundant, export more skill-intensive value-added. This is true for different specifications and both for the cross-sectional data as well as the panel. However, using gross exports instead of value-added, the authors find no support for the prediction of the model with respect to manufacturing. \par 
Both for services and total trade, value-added coefficients also lean more towards the predictions of the Heckscher-Ohlin Theorem. However, the difference to the gross trade coefficients is not significant. Applying robustness checks developed by \citeA{Romalis} and \citeA{Trefler.2010}, they find further confirmation for their previously obtained results. The main contribution of \citeA{Ito.2017} is twofold: First, they demonstrate that using value-added and gross export data may not always lead to the same result. Second, and more importantly, they show that, in the context of skill-abundance, the Heckscher-Ohlin Theorem finds stronger support with value-added data, although only for manufacturing.

\section{Data \& Methodology}
\paragraph{Data.} The empirical analysis is based on the two latest releases, i.e. 2013 and 2016, of the World Input-Output Database as well as the respective Socioeconomic Accounts \cite{Timmer.2015}.\footnote{~The WIOD is one of three major databases for inter-country Input-Output Tables. The remaining two databases are OECD TiVA and EORA. We decided to use the WIOD for two reasons: (1) Data quality: The EORA would be the most attractive in terms of coverage, but there has been severe critique with respect to data quality \cite[p.~155]{Kowalski.2015}. (2) Industry Classification: In comparison to the OECD TiVA, the WIOD uses a more detailed industry classification scheme.} The 2016 release is applied for calculating value-added exports. It covers 56 industries in 43 different countries plus the rest of the world, annually between 2000 and 2014. A comprehensive list of countries is provided in the Appendix. The industries are classified according to the International Standard Industrial Classification (ISIC) Rev. 4. \par
Value-added exports can be retrieved from inter-country Input-Output Tables following \citeA{Johnson.2012} and \citeA{Koopman.2014}. Also, \citeA{Aslam.2017} provide an accessible summary of the method. Consider the basic structure of a general Input-Output-Table with $S$ industries and $C$ countries and thus a total number of $N=S \times C$ observed industries, where the gross output vector $\textbf{X} \in \mathbb{R}^{N}$ is given by 
\begin{equation}
    \textbf{X}=(\textbf{I}-\textbf{A})^{-1}\textbf{Y}
\end{equation}
The vector of final demand is denoted as $\textbf{Y} \in \mathbb{R}^{N\times C}$, while $\textbf{A} \in \mathbb{R}^{N\times N}$ is the matrix of input-output-coefficients, where $[\textbf{A}]_{ij}$ indicates how many units of $x_i$ are required to produce one unit of $x_j$. The matrix $(\textbf{I}-\textbf{A})^{-1}$ is the well-known Leontief inverse with $[(\textbf{I}-\textbf{A})^{-1}]_{ij}$ reflecting how many units of $x_i$ are required to produce one unit of the final good $y_j$. \par
Then, we define the diagonal matrix $\textbf{V} \in \mathbb{R}^{N\times N}$, which captures the domestic value-added shares, such that
\begin{equation}
    [\textbf{V}]_{i}=[\textbf{I}]_{i}-\sum_{n=1}^{N} [\textbf{A}]_{ni}
\end{equation}
Value-added exports of industry $i$ in country $c$ to country $d \neq c$ are now given as an element of the matrix $\textbf{VX} \in \mathbb{R}^{N\times C}$, which is defined as
\begin{equation}
    \textbf{VX} = \textbf{V} \times (\textbf{I}-\textbf{A})^{-1} \times \textbf{Y}
\end{equation}
Eventually, value-added exports for 2,290 industries (of 2,264 available in total) to 42 countries each are observed. This is, because rows and columns containing only zeros need to be excluded, since singularity issues in obtaining the Leontief inverse otherwise occur.  \par
Then, we reshape the matrix $\textbf{VX}$ for each year and create one balanced panel dataset including 1,476,420 observations, where the value-added exports of each industry in all covered countries to each other country in the dataset are recorded. \par
Moreover, the Socioeconomic Accounts, which are provided in the WIOD, are included in our analysis. Specifically, this includes industry-level data on the capital stock, the number of employees, and capital and labor compensation. This can be utilized to define a labor-capital ratio at industry-level. These data are matched to the value-added export data. \par
Additionally, we include data of the Socioeconomic Accounts of the 2013 WIOD-release, since these include data on educational attainment, i.e. the hours worked by low-, medium-, and high-skilled employees on industry level. This is the data used by \citeA{Ito.2017}. However, we face two problems in using this data. First, the older release covers fewer countries, and different years (1995 to 2009). Second, it is based on a different industry classification system, i.e. ISIC Rev. 3. Nonetheless, we impute the data on skill-levels for the countries and years we have by matching the previous industry classification to the new one. All in all, the number of observations with respect to educational attainment is significantly smaller than for capital and labor. We will, thus, include regression models containing skill-levels include rather as a robustness check and replication exercise with respect to \citeA{Ito.2017}. 

\paragraph{Methodology.} Methodologically, we closely follow \citeA{Ito.2017}. Given the available data, we employ two different definitions of labor-capital intensity and endowments. That is, \textit{(i)} as the ratio between total labor compensation and capital compensation in the respective industry, and \textit{(ii)} as the ratio between total hours worked and the nominal capital stock. \par
To test for the Heckscher-Ohlin Theorem, we estimate the following model
\begin{equation}
\label{eq:regressionmodel}
    log(VX)_{odit} = \alpha_{dit} + \sigma_{ot} + \beta_1 log\Bigl(\frac{l}{k}\Bigl)_{oit} + \beta_2 \Bigl[log\Bigl(\frac{l}{k}\Bigl)_{oit} \times log\Bigl(\frac{L}{K}\Bigl)_{ot}\Bigl] + \varepsilon_{odit}
\end{equation}
The subscripts $o$, $d$, $i$, and $t$ denote the exporting country, the destination country, the exporting industry, and time, respectively. That is, $\alpha_{dit}$ and $\sigma_{ot}$ are importer-industry-year- and exporter-year-dummies. Lower-case letters denote industry-level variables, while upper-case letters denote country-level ones. The coefficient of interest is $\beta_2$. For a given exporting country and a given market, which is captured in $\alpha_{dit}$, a labor-abundant country should engage in value-added exports of labor-intensive goods. Hence, a significantly positive $\beta_2$ can be seen as supporting the Heckscher-Ohlin model. \par
In robustness checks, we include data on skill intensity and endowments. Analogously to \citeA{Ito.2017}, we combine medium- and low-skilled employees to a group of unskilled labor force. Hence, skill intensity is measured as the ratio between the hours worked by high-skilled and unskilled employees. The extended regression model, then, is
\begin{multline}
\label{eq:regressionmodel_extended}
    log(VX)_{odit} = \alpha_{dit} + \sigma_{ot} + \beta_1 log\Bigl(\frac{l}{k}\Bigl)_{oit} + \beta_2 \Bigl[log\Bigl(\frac{l}{k}\Bigl)_{oit} \times log\Bigl(\frac{L}{K}\Bigl)_{ot}\Bigl] \\ + \gamma_1 log\Bigl(\frac{l_{hs}}{l_{us}}\Bigl)_{oit} + \gamma_2 \Bigl[log\Bigl(\frac{l_{hs}}{l_{us}}\Bigl)_{oit} \times log\Bigl(\frac{L_{hs}}{L_{us}}\Bigl)_{ot}\Bigl] + \varepsilon_{odit}
\end{multline}

\section{Results}
The regression results are displayed in Table \ref{tab:results1}. Columns 1 and 2 show the estimation results for labor-capital intensity and endowments defined as the ratio of factor compensations, while columns 3 and 4 refer to the definition of the ratio of hours and the nominal capital stock. For all regressions we use robust standard errors accounting for heteroskedasticity \cite{White.1980}. \par
Defining labor-capital intensity and endowments as the ratio of factor compensations, we find a significantly positive coefficient for $\beta_2$ (column 1). This suggests that countries with an abundance in labor export relatively more goods, which use labor more intensively. This is in line with the prediction of the Heckscher-Ohlin Theorem.  \par
However, when we define the labor-capital intensities and endowments as the ratio of hours worked to total capital stock, our results change (column 3). Coefficients for this specification can be seen in column 3, where $\beta_2$ is significantly negative. This would suggest that labor-abundant countries export relatively fewer goods which use labor intensively in their production. \par
As a robustness check, we perform the same regression as \citeA{Ito.2017} and include terms that capture the relative skill abundances and intensities. The results are shown in columns 2 and 4 of Table \ref{tab:results1}. Our sample becomes significantly smaller due to the restricted data availability, as outlined in Chapter 4, but the coefficients $\beta_1$ and $\beta_2$ are similar to the first model. Results with respect to skill-abundance and intensity are in line with what \citeA{Ito.2017} report. There, the authors also find positive and significant coefficients for skill intensities and endowments. However, in their case, coefficients capturing relative labor endowments and intensities are insignificant, contrary to what we find. \par
A possible explanation for the contradictory results between the two applied definitions of labor-capital intensity and endowments lies in the Leontief paradox \cite{Leontief}. One crucial assumption of the Heckscher-Ohlin model is that countries have identical production technologies. \citeA{Leontief} did not account for such differences in technologies when he tested the Theorem and, as a result, found evidence against it. Only later, when this assumption was relaxed and differences in technologies where being accounted for, empirical support for the Theorem emerged. This is reflected in our results as well. When we use the ratio of hours worked to the capital stock, our results contradict the Theorem. This is due to the fact that this specification does not account for differences in technologies between countries. However, when we define the labor-capital intensities and endowments as the ratio of compensations, we implicitly control for differences in technologies. As a result, we find support for the Heckscher-Ohlin Theorem. \par

\begin{table}[!h] \centering 
  \caption{Regression results} 
  \label{tab:results1} 
    \footnotesize
\begin{tabular}{@{\extracolsep{5pt}}lcccc} 
\\[-1.8ex]\hline 
\hline \\[-1.8ex] 
 & \multicolumn{4}{c}{\textit{Dependent variable:}} \\ 
\cline{2-5} 
\\[-1.8ex] & \multicolumn{4}{c}{log(VX)} \\ 
\\[-1.8ex] & \multicolumn{2}{c}{Compensations} & \multicolumn{2}{c}{Hours/capital stock}\\
\\[-1.8ex] & (1) & (2) & (3) & (4)\\ 
\hline \\[-1.8ex] 
 $log\bigl(\frac{l}{k}\bigl)$ & $-$0.324$^{***}$ & $-$0.358$^{***}$ & $-$0.240$^{***}$ & $-$0.235$^{***}$  \\ 
  & (0.002) & (0.003) & (0.003) & (0.004) \\ 
  & & & & \\ 
$log\bigl(\frac{l}{k}\bigl) \times log\bigl(\frac{L}{K}\bigl)$ & 0.245$^{***}$ & 0.254$^{***}$ & $-$0.015$^{***}$ & $-$0.012$^{***}$ \\ 
  & (0.004) & (0.006) & (0.0004) & (0.001) \\ 
  & & & & \\ 
 $log\bigl(\frac{l_{hs}}{l_{us}}\bigl)$ &  & 0.201$^{***}$ &  & 0.115$^{***}$ \\ 
  &  & (0.008) &  & (0.008) \\ 
  & & & & \\ 
 $log\bigl(\frac{l_{hs}}{l_{us}}\bigl) \times log\bigl(\frac{L_{hs}}{L_{us}}\bigl)$ &  & 0.156$^{***}$ &  & 0.093$^{***}$ \\ 
  &  & (0.004) &  & (0.004) \\ 
  & & & & \\ 
\hline \\[-1.8ex] 
Observations & 1,353,446 & 734,464 & 1,409,088 & 763,016 \\ 
R$^{2}$ & 0.161 & 0.158 & 0.132 & 0.122 \\ 
Adjusted R$^{2}$ & 0.161 & 0.158 & 0.132 & 0.122 \\ 
%F Statistic & 16,216.250$^{***}$ (df = 16; 1353429) & 10,573.650$^{***}$ (df = 13; 734450) & 13,409.070$^{***}$ (df = 16; 1409071) & 8,141.752$^{***}$ (df = 13; 763002) \\ 
\hline 
\hline \\[-1.8ex]
\multicolumn{5}{l}{\textit{Note:} Standard errors accounting for heteroskedasticity \cite{White.1980}  } \\
\multicolumn{5}{l}{are applied. $^{*}$p$<$0.1, $^{**}$p$<$0.05, $^{***}$p$<$0.01} \\ 
\end{tabular} 
\end{table}

\begin{table}[!htbp] \centering 
  \caption{Regression results per broad industry category according to ISIC Rev. 4} 
  \label{tab:results2} 
  \footnotesize
\begin{tabular}{@{\extracolsep{5pt}}lcccccc} 
\\[-1.8ex]\hline 
\hline \\[-1.8ex] 
 & \multicolumn{6}{c}{\textit{Dependent variable:}} \\ 
\cline{2-7} 
\\[-1.8ex] & \multicolumn{6}{c}{log(VX)} \\ 
\\[-1.8ex] & A & B & C & D,E & F & G\\ 
\hline \\[-1.8ex] 
 $log\bigl(\frac{l}{k}\bigl)$ & $-$0.046$^{***}$ & $-$1.018$^{***}$ & $-$0.437$^{***}$ & $-$0.122$^{***}$ & $-$0.404$^{***}$ & $-$0.207$^{***}$ \\ 
  & (0.005) & (0.014) & (0.004) & (0.010) & (0.012) & (0.009) \\ 
  & & & & & & \\ 
 $log\bigl(\frac{l}{k}\bigl) \times log\bigl(\frac{L}{K}\bigl)$ & $-$0.319$^{***}$ & 1.055$^{***}$ & 0.109$^{***}$ & 0.337$^{***}$ & 0.583$^{***}$ & $-$0.414$^{***}$ \\ 
  & (0.013) & (0.021) & (0.008) & (0.018) & (0.027) & (0.022) \\ 
  & & & & & & \\ 
\hline \\[-1.8ex] 
Observations & 68,026 & 26,574 & 491,791 & 76,003 & 25,972 & 78,776 \\ 
R$^{2}$ & 0.091 & 0.294 & 0.165 & 0.166 & 0.211 & 0.232 \\ 
Adjusted R$^{2}$ & 0.091 & 0.294 & 0.165 & 0.165 & 0.211 & 0.232 \\ 
%F Statistic & 426.000$^{***}$ (df = 16; 68009) & 691.960$^{***}$ (df = 16; 26557) & 6,094.980$^{***}$ (df = 16; 491774) & 942.466$^{***}$ (df = 16; 75986) & 434.676$^{***}$ (df = 16; 25955) & 1,488.457$^{***}$ (df = 16; 78759) \\ 
\hline 
\hline \\[-1.8ex] 
%\textit{Note:}  & \multicolumn{6}{r}{$^{*}$p$<$0.1; $^{**}$p$<$0.05; $^{***}$p$<$0.01} \\ 
\end{tabular} 
\begin{tabular}{@{\extracolsep{5pt}}lcccccc} 
%\\[-1.8ex]\hline 
%\hline \\[-1.8ex] 
% & \multicolumn{6}{c}{\textit{Dependent variable:}} \\ 
%\cline{2-7} 
%\\[-1.8ex] & \multicolumn{6}{c}{log(VX)} \\ 
\\[-1.8ex] & H & I & J & K & L & M\\ 
\hline \\[-1.8ex] 
$log\bigl(\frac{l}{k}\bigl)$ & $-$0.538$^{***}$ & $-$0.146$^{***}$ & $-$0.149$^{***}$ & $-$0.213$^{***}$ & $-$0.419$^{***}$ & $-$0.072$^{***}$ \\ 
  & (0.006) & (0.018) & (0.007) & (0.008) & (0.006) & (0.007) \\ 
  & & & & & & \\ 
$log\bigl(\frac{l}{k}\bigl) \times log\bigl(\frac{L}{K}\bigl)$ & 0.647$^{***}$ & 0.249$^{***}$ & $-$0.056$^{***}$ & 0.158$^{***}$ & 0.030$^{***}$ & 0.060$^{***}$ \\ 
  & (0.011) & (0.047) & (0.014) & (0.016) & (0.006) & (0.016) \\ 
  & & & & & & \\ 
\hline \\[-1.8ex] 
Observations & 123,840 & 26,402 & 100,018 & 70,004 & 26,445 & 108,016 \\ 
R$^{2}$ & 0.234 & 0.179 & 0.175 & 0.145 & 0.335 & 0.138 \\ 
Adjusted R$^{2}$ & 0.234 & 0.178 & 0.175 & 0.145 & 0.334 & 0.138 \\ 

\hline 
\hline \\[-1.8ex] 
\multicolumn{7}{l}{\textit{Note:} Standard errors account for heteroskedasticity \cite{White.1980}. $^{*}$p$<$0.1, $^{**}$p$<$0.05, $^{***}$p$<$0.01} \\
\multicolumn{7}{l}{For a list of industry codes see page 43 of the \href{https://unstats.un.org/unsd/publication/seriesM/seriesm_4rev4e.pdf}{ISIC documentation}.}
\end{tabular} 
\end{table} 

In Table \ref{tab:results2} we report results for each broad industries according to the International Standard Industrial Classification (ISIC).\footnote{For a list of industry codes see page 43 of the \href{https://unstats.un.org/unsd/publication/seriesM/seriesm_4rev4e.pdf}{ISIC documentation}.} Endowments and intensities are defined as the ratio of factor compensations. While all coefficients are significant, we do not find universal support for Heckscher-Ohlin. That is, for nine out of twelve industries, we are able to report a positive coefficient $\beta_2$. Among those are e.g. construction, manufacturing and energy supplies. For wholesale and resale trade, agriculture and information and communication we find evidence against the Theorem. \par
Lastly, we report results for selected years, i.e. 2000, 2007, and 2014, in Table \ref{tab:results3}, defining intensities and endowments as the ratio of factor compensations. It is apparent that the explanatory power of the Heckscher-Ohlin model slightly decreases over time, especially between 2000 and 2007. The adjusted R$^2$ decreases from 0.377 in 2000 to 0.298 in 2007. One potential explanation for this observation is that trade explained by Krugman's trade model \cite<see>{krugman} became more relevant.

\begin{table}[!htbp] 
\centering 
  \caption{Results for selected years} 
  \label{tab:results3} 
\begin{tabular}{@{\extracolsep{5pt}}lccc} 
\\[-1.8ex]\hline 
\hline \\[-1.8ex] 
 & \multicolumn{3}{c}{\textit{Dependent variable:}} \\ 
\cline{2-4} 
\\[-1.8ex] & \multicolumn{3}{c}{log(VX)} \\ 
\\[-1.8ex] & 2000 & 2007 & 2014 \\ 
\hline \\[-1.8ex] 
 $log\bigl(\frac{l}{k}\bigl)$ & $-$0.293$^{***}$ & $-$0.291$^{***}$ & $-$0.349$^{***}$  \\ 
  & (0.015) & (0.015) & (0.015) \\ 
  & & &  \\ 
 $log\bigl(\frac{l}{k}\bigl) \times log\bigl(\frac{L}{K}\bigl)$ & 0.235$^{***}$ & 0.214$^{***}$ & 0.267$^{***}$   \\ 
  & (0.029) & (0.031) & (0.032)  \\ 
  & & &  \\ 
\hline \\[-1.8ex] 
Observations & 91,256 & 93,148 & 92,312  \\ 
R$^{2}$ & 0.393 & 0.317 & 0.315 \\ 
Adjusted R$^{2}$ & 0.377 & 0.298 & 0.296  \\ 
\hline 
\hline \\[-1.8ex] 
\textit{Note:}  & \multicolumn{3}{r}{$^{*}$p$<$0.1; $^{**}$p$<$0.05; $^{***}$p$<$0.01} \\ 
\end{tabular} 
\end{table}

\section{Conclusion}
Ever since its publication, economists have tried to find empirical validation for the Heckscher-Ohlin Theorem. The majority of the earlier attempts, among them the famous Leontief Paradox \cite{Leontief}, were unable to find support for it. After \citeA{Trefler} proposed that the assumption of identical technologies might not hold, economists started taking this into account. However, the evidence was still far from conclusive. One possible reason for this is the gross trade data that was applied. It was first pointed out by \citeA{Leamer} that for providing evidence in favor of Hechscker-Ohlin value-added trade data may be necessary. The only recent emergence of such data has enabled the testing for the Theorem with value-added trade data. One recent paper by \citeA{Ito.2017} test for the Heckscher-Ohlin Theorem with value-added and gross trade data with respect to skill-abundance and skill-intensity, finding that value-added trade data more strongly support the predictions for manufacturing. \par
In our paper we follow the approach of \citeA{Ito.2017}. However, we use a more recent release of the World Input-Output Database that includes more countries and a more detailed industry classification. We define two different measurements of labor-capital intensities and endowments. Defining it as the ratio between hours worked and the capital stock, we do not find support for the Heckscher-Ohlin Theorem. However, defining intensities and endowments as the ratio of factor compensations, results strongly support it. A likely explanation for this is that the latter definition takes differences in technologies into account and leads to support for the Theorem. We also compare different industries and find that in nine out of twelve cases, the Heckscher-Ohlin model can explain trade flows. 

\bibliographystyle{apacite}
\bibliography{intecon.bib}

\section*{Appendix}
\paragraph{List of countries (ISO 3166-1).} AUS, AUT, BEL, BGR, BRA, CAN, CHE, CHN, CYP, CZE, DEU, DNK, ESP, EST, FIN, FRA, GBR, GRC, HRV, HUN, IDN, IND, IRL, ITA, JPN, KOR, LTU, LUX, LVA, MEX, MLT, NLD, NOR, POL, PRT, ROU, RUS, SVK, SVN, SWE, TUR, TWN, USA.

\end{document}